\documentclass[a4paper]{jpconf}
\usepackage{graphicx}
\begin{document}
\title{Interplay of superconductivity and magnetism in a $t$-$t'$-$J$ approach to high $T_c$ cuprates }

\author{P A Marchetti}

\address{Dipartimento di Fisica e Astronomia, U. di Padova 
 and INFN, I-35131 Padova, Italy}

\ead{marchetti@pd.infn.it}

\begin{abstract}
 We review a recently proposed mechanism for superconductivity in hole-doped cuprates exhibiting a strong interplay between pairing and antiferromagnetism. Starting from the $t$-$t'$-$J$ model for the CuO planes, we show that this interplay can explain in a unified framework the pseudogap phenomenology of the spectral weight of the hole, the hourglass-like structure of the magnetic excitation, the critical exponent of the superfluid density, the relation between the scale of the magnetic resonance and $T_c$. 
 
\end{abstract}

\section{Common origin of short-range AF and pairing}
In this note we outline the basic ideas of our approach to the low-energy physics of hole-doped cuprates \cite{jcmp},\cite{mysy},\cite{mg}, emphasizing an explicit interplay between antiferromagnetism (AF) and pairing attraction leading to superconductivity (SC).

 We tackle the non-perturbative constraint of no-double occupation of the  $t$-$t'$-$J$ model
by decomposing at each site $i$ the fermionic spin 1/2 hole field $c_\alpha$ into a product of a bosonic chargeless spin 1/2 boson $b_\alpha$, the spinon, and a charged spinless fermion $h$, the holon: $c_{\alpha i}=b_{\alpha i} h_i$. Then the Pauli exclusion principle for the holon enforces automatically the constraint.  However this decomposition involves an unphysical degree of freedom, because one can multiply at each site the spinon and the holon by arbitrary opposite phase factors leaving the physical hole field unchanged. This local $U(1)$ gauge invariance can be made manifest by introducing a slave-particle gauge field $A_\mu$. The gauge field induces an attraction between holon and spinon with $T$-dependent damping. The spin-charge decomposition is a useful approach only if this attraction does not lead to confinement, which in fact doesn't occur in the considered situations.
In planar (2D) and in linear (1D) systems, we have an additional option to improve a Mean Field (MF) treatment: gauging a global symmetry of the system one can bind statistical fluxes to
particle-fields, resulting only in  a change of their
statistics.  The introduction of these
fluxes in the lagrangian formalism is realized through statistical
Chern-Simons gauge fields.  In our case, the hole carries both charge
and spin degrees of freedom, so we can bind a statistical charge flux $\Phi^h$ to the holon and a statistical spin flux $\Phi^s$ to the spinon. Assuming the absolute value of the spin flux to be 1/2 and the charge flux to be -1/2 the resulting  hole field $c_{\alpha }=(\exp{(i\Phi^s)}b)_{\alpha} \exp{(i\Phi^h)} h$  is still fermionic. With this choice the holon bound to the charge flux and the spinon bound to the spin flux obey semion statistics, implying that they acquire under exchange a phase $\pm i$, intermediate between the bosonic $+1$ and the fermionic $-1$, hence the name ``semion''. These semionic holons are expected to obey Haldane statistics of order 2 in momentum space, meaning that a maximum of two semions are allowed to have the same momenta. Hence a gas of spinless semions of finite density has a (pseudo-)Fermi surface at low $T$ coinciding with that of spin 1/2 fermions with the same density.
 
If no approximations are made, all the slave-particle approaches to the $t$-$t'$-$J$ model are strictly equivalent. However, as soon as MF approximations are made, this approach is a new slave-semion approach, distinct
from  slave boson or slave fermion approaches. The semionic statistics is the same found in the solution of the 1D model \cite{hmsy}.
In the improved semionic mean field
approximation (MFA),  the spinon configurations in the presence of moving holons are first
optimized, leading to a new bosonic spinon denoted by $s$, describing fluctuations around the optimized spinon background and still satisfying the constraint
$s^{*}_{i\alpha}s_{i\alpha}=1$ (summation over repeated spin indices is understood).  From now on it is this spinon that we refer to.
In the adopted MFA we neglect the
holon fluctuations in $\Phi^h$ and the spinon fluctuations in
$\Phi^s$ . This leads to a much simpler form of the two statistical
fluxes. The charge one  is actually static and it provides a $\pi$-flux phase factor per
plaquette. This flux converts, via Hofstadter mechanism,  the low-energy modes of the spinless holons $h$ into Dirac fermions with dispersion defined in the Magnetic Brillouin Zone
(MBZ) and a small Fermi Surface (FS) $\epsilon_F \sim
t\delta$, where $\delta$ is the doping concentration, characterizing the "pseudogap phase" (PG) of the model. Increasing doping or temperature one reaches a crossover line $T^*$, identified with the experimental inflection point of in-plane resistivity. Crossing this line we enter in the "strange metal phase"(SM)  in which  the effect of the charge flux is screened by the optimal $b$-spinon configuration in MFA and we recover a ``large'' FS for the holons with $\epsilon_F \sim t (1+\delta)$.
For the SU(2) spin flux in MFA only
the $\sigma_z$-component survives:
\begin{eqnarray}
\label{eq:5}
\Phi^s(x) = \sigma_z \sum_l
{h}^{*}_l {h}_l\frac{(-1)^{|l|}}{2} 
\arg(\vec{x}-\vec{l}),
\end{eqnarray}
and its gradient can be viewed as the potential of spin-vortices with the axis along the direction of the AF background.  These vortices appear in the U(1) subgroup of the spin group complementary to the coset  labeling the directions of the spin. Fluctuations of such directions describe the spin-waves, viewed as composites of spinons generated by gauge attraction between spinon and antispinon. Therefore the spin-vortices have a purely quantum origin, somewhat analogous to the Aharonov-Bohm effect. Note that there
is the holon density  in the right hand side of
Eq.(\ref{eq:5}), so that the spin-vortices are always
centered on the holes. The vorticity or chirality of the vortices
is $(-1)^{|l|}$, with
$|l|=l_x+l_y$. The effect of the optimal spin flux is then to
attach a spin-vortex to the holon, with opposite chirality on the
two N\'eel sublattices. 
These vortices take into account
the long-range quantum distortion of the AF background caused by
the insertion of a dopant hole. They play also a crucial role in development of superconductivity and their chirality structure is a consequence of antiferromagnetic interaction, showing a first manifestation of the interplay between SC and magnetism in this approach.

The optimization of the $b$-spinons discussed above involves also a spin-flip associated to every holon jump between different N\'eel sublattices. Therefore in the $t,t'$-terms
the spinons appear in the ``ferromagnetic'' Affleck-Marston
 form $AM_{\langle ij \rangle} =(s_i
e^{-i\Phi^s_i})_{\alpha}(e^{i\Phi^s_j}s_{j})_{\alpha}$, whereas in the $J$-term they appear in the ``antiferromagnetic'' RVB form
$RVB_{\langle ij \rangle}= \epsilon^{\alpha\beta} (e^{i\Phi^s_i}s_i)_{\alpha}(e^{i\Phi^s_j}s_{j})_{\beta}$.
The above AM/RVB dichotomy is characteristic of the slave-semion approach
involving the SU(2) spin rotation group even in 1D, where it has
been rigorously derived \cite{hmsy}. It has the appealing feature that
 optimizing the hopping $(|AM_{\langle ij \rangle}|=1)$ we optimize also the Heisenberg term. In fact, neglecting $A$-fluctuations, the leading terms of the
Hamiltonian can be written as:
\begin{eqnarray}
  \label{tJ} &\;H= \sum_{nn \langle ij \rangle} (-t) {AM}_{ij}
  h^{*}_i h_j e^{i (\Phi^h_i-\Phi^h_j) } + h.c. +\\
  &\;  \sum_{nnn \langle ij \rangle} (t') {AM}_{ij}
  h^{*}_i h_j  + h.c. +\nonumber \\
  &\; \sum_{nn \langle ij \rangle} J(1-h^*_ih_i-h^*_jh_j) (1-|{AM}_{{\langle ij \rangle}}|^2)
  + J h^*_ih^*_jh_jh_i |{RVB}_{\langle ij \rangle}|^2, \nonumber
\end{eqnarray}
where in the second to last term we used the identity holding for a bosonic spinon $ |AM_{\langle ij \rangle}|^2 + |RVB_{\langle ij \rangle}|^2 =1$.
A long-wavelength treatment of this term, neglecting the spin-vortices, leads to a
CP$^1$ spinon nonlinear $\sigma-$model. 

The additional interaction term between spinons and spin-vortices is then of the form
\begin{equation}
\label{spvo}
J (1-2 \delta)(\nabla{\Phi^s(x)})^2  s^* s(x)
\end{equation}
and it is the source of both short-range AF and charge pairing, providing a second interplay between SC and AF. In fact, from a quenched treatment of spin-vortices we derive the MF expectation value
$\langle(\nabla{\Phi^s(x)})^2\rangle = m_s^2 \approx 0.5\delta |\log
\delta|$, which opens a mass gap for the spinon, consistent with AF
correlation length at small $\delta$ extracted from the neutron
experiments \cite{ke}.
 Thus, propagating in the gas of slowly moving spin-vortices, the AF spinons, originally gapless in the undoped Heisenberg model,
acquire a finite gap, leading to a short range AF order.
By averaging instead the spinons in Eq. (\ref{spvo}), we obtain an effective interaction:
 \begin{equation}
\label{zh}
J (1-2 \delta) \langle s^* s \rangle \sum_{i,j} (-1)^{|i|+|j|} \Delta^{-1}
 (i - j) h^*_ih_i h^*_jh_j,
\end{equation}
 where $\Delta$ is the 2D lattice laplacian.
From Eq. (\ref{zh}) we see that the interaction mediated by spin-vortices on holons is of 2D Coulomb type. From the known behaviour of planar Coulomb systems we derive that below a crossover temperature
$T_{ph} \approx J (1-2 \delta)  \langle s^* s \rangle $ , which turns out to be greater than $T^*$, a finite density of incoherent holon pairs appears.
Therefore the origin of the charge-pairing is magnetic, but it is not due to exchange of AF spin fluctuations.
  
More in detail, if we describe the magnetic Brillouin Zone (BZ) with the upper half of the BZ, in PG the BCS treatment of this pairing on the two small FS centered at $(\pm \pi/2, \pi/2)$ yields 2 p-wave orders for the holon, recombining to give a d-wave order in the full BZ for the hole, as first suggested in a different setting in Ref. \cite{sus}. Since the pairing distinguishes the two N\'eel sublattices, in SM (but below $T_{ph}$) it produces also a folding of the holon FS into the magnetic BZ, inducing  the formation of two hole-like FS around $(\pm \pi/2, \pi/2)$ and an electron-like FS around  $(\pm \pi, 0) = (0, \pi)$. In BCS approximation we have again p-wave order in the two hole-like FS and s-wave order in the electron-like FS for the holon, reproducing finally a d-wave order for the hole in the full BZ \cite{mg}. The pairing turns out to be not far from the BCS-BEC crossover, but still in the BCS side. Since the pairing originates from spin-vortices it is independent  of  nesting features of the Fermi surface, used in most spin-wave approaches. 

\section{Consequences of the interplay between magnetism and pairing}
However we do not have condensation of holon pairs because the fluctuations of the phase of the pairing field are too strong.
In fact the scattering of the phase of the holon-pair field, with a gap $\sim T$, against holons destroys the holon-pair coherence of BCS approximation. Via spinon-holon binding it produces the phenomenology of  Fermi arcs coexisting with gap in the antinodal region \cite{no}. I.e. lowering T one finds a decrease of the hole spectral weight on the FS and the formation of superconducting-like peaks even in the normal state, starting from the antinodal region, but with a strong overdamping due to the strong interaction of the spinon with gauge fluctuations. The final outcome \cite{mg} agrees qualitatively with experimental data on tunneling, ARPES and conductivity.  The aspects of pseudogap phenomenology related to the gradual reduction of the spectral weight are therefore a consequence of the AF responsible for the pairing.

This charge pairing does not yet lead to hole-pairing, since the spins are still unpaired. The hole-pairing is achieved by the gauge attraction between holon and spinon which, using the holon-pairs as sources of attraction,  induces in turn the
formation of short-range spin-singlet (RVB) spinon pairs, see Fig. 1 (left).
\begin{figure}[ht]
\includegraphics[width=21pc]{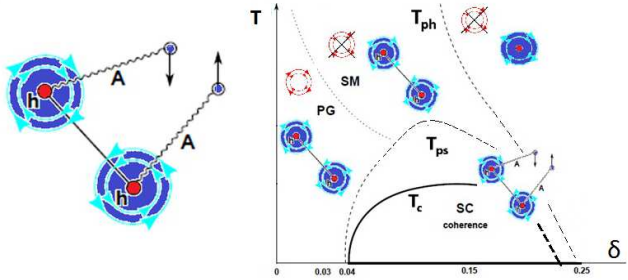}\hspace{2pc}%
\begin{minipage}[b]{14pc}\caption{\label{Fig.1}Left: pictorial representation of hole pairs: holons
(red circles) surrounded by spin-vortices (blue), spinons (blue circles with spin-arrow), spin-vortex attraction (black line), gauge attraction (wavy line). Right: pictorial representation of the crossovers discussed in the text: charge-flux as charge vortex (in red).}
\end{minipage}
\end{figure} 
 This phenomenon occurs, however, only when the density of holon-pairs is sufficiently high, since this attraction has to overcome
the original AF-repulsion of spinons caused by the Heisenberg $J$-term which is positive in the slave-semion approach  (see Eq. (\ref{tJ})). Summarizing, at an intermediate crossover temperature  lower than $T_{ph}$ and denoted by $T_{ps}$, a finite density of incoherent spinon RVB pairs appears; combined with the holon pairs it gives rise to a gas of incoherent preformed hole pairs. The lowering of free energy allowing the formation of spinon pairs is due to the fact that  the spinon gap originates from screening due to unpaired vortices, the short-range vortex-antivortex  pairs essentially not contributing  to it. Therefore the spinon gap lowers proportionally to the density of spinon pairs, implying a lowering of the kinetic energy of spinons, a mechanism definitely not BCS-like. More precisely, for a finite density of spinon pairs there are two (positive energy) spinon excitations, with different energies, but the
same spin and momenta. For example for spin up they are obtained by creating a spinon up unpaired or destructing  a spinon down in one of the RVB pairs. The corresponding dispersion relation, thus exhibits two
(positive) branches:
\begin{equation}
\label{sd} \omega (\vec k) =  \sqrt{(m_s^2 - |\Delta^s|^2) +
(|\vec k| \pm |\Delta^s|)^2,}
\end{equation}
where $\Delta^s$ is (proportional to) the RVB spinon pair order parameter.
 The lower branch exhibits a minimum with an  energy lower than $m_s$ corresponding to the reduction of the spinon gap discussed above, and it implies a backflow of the gas of spinon-pairs dressing the ``bare'' spinon.  The spinon-antispinon attraction mediated by gauge
fluctuations induces a similar structure for the magnon dispersion around the AF wave vector \cite{gam}, see Fig.2, reminiscent of the hourglass shape of spectrum found in neutron experiments \cite{hour}, with an energy at the AF wave vector approximately twice the spinon gap $\approx J (1-2\delta) |\delta\ln\delta|^{1/2}$.

\begin{figure}[ht]
\includegraphics[width=15pc]{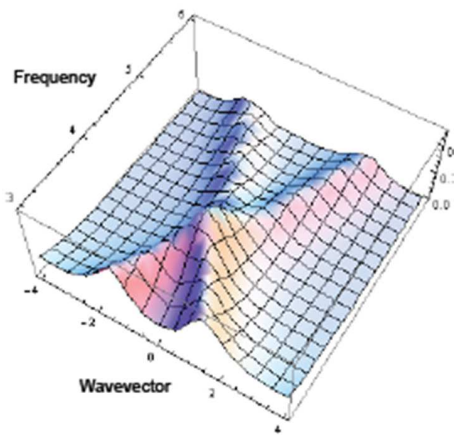}\hspace{2pc}%
\begin{minipage}[b]{14pc}\caption{Plot (in arbitrary units) of the imaginary part of the magnon retarded propagator in SC phase derived as discussed in the text \cite{gam}. Wavevectors are measured starting from $(\pi,\pi)$.}
\label{Fig.2}
\end{minipage}
\end{figure}

 This feature is evident in the superconducting phase, where the damping induced by gauge fluctuations in the normal phase is strongly suppressed. Hence the modification of the magnetic excitation leading to the magnetic resonance is ultimately due to the charge pairing interaction.

The gradient of the phase of the low-energy treatment of the RVB-singlet
hole-pair field 
$h_i^*h_j^* \epsilon^{\alpha\beta} (s_i
e^{i\Phi^s_i})_{\alpha}(e^{i\Phi^s_j}s_{j})_{\beta}$ 
describes magnetic vortices (not to be confused with quantum spin-vortices). Between $T_{ps}$ and the superconducting transition these magnetic vortices are in the plasma phase, because the hole pairs are incoherent.
Therefore we propose to identify $T_{ps}$  with the experimental crossover
corresponding to the appearance of the diamagnetic and (vortex)
Nernst signal \cite{ong}. This interpretation is reinforced by the
computation of the equi-level plot of the spinon pair
 density in the $\delta$-$T$ plane \cite{mysy}, resembling the
 contour-plot of the Nernst/diamagnetic signal. The doping density enters in the above mechanism through two factors:
 the density of hole pairs and the strength of the attraction behaving like
 $\approx J(1-2\delta)$ from eq.(\ref{zh}). These two effects act in an opposite way
 increasing doping, thus yielding a ``dome'' shape to $T_{ps}(\delta)$, starting from a non-zero doping concentration. 

Superconductivity occurs by condensation of hole pairs at a temperature $T_c$ lower than $T_{ps}$
inheriting the dome structure, see Fig.1(right).

 Below $T_{ps}$ the effective action obtained integrating out holons and spinons is a $A$-gauged 3D XY model, where the angle-field of the XY model is the phase of the hole-pair field. At the SC transition the gauge field $A$ is gapped by the Anderson-Higgs mechanism and the gauge fluctuations are suppressed. Hence the superconducting transition is almost of classical 3D XY type, driven by condensation of magnetic vortices related to phase coherence as in BEC systems and providing another interplay between SC and magnetism. The transition is therefore not BCS-like, as shown by the critical exponent of the superfluid density which is 2/3 \cite{mb}, characteristic of the 3D XY model. This value turns out to agree with the experimental data in underdoped-optimally doped cuprates \cite{kam}.

Finally, the common origin of pairing and short-range AF from
 the same term in the representation of the $t$-$t'$-$J$ model Eq.(\ref{spvo}) produces an intrinsic approximately linear relation between the energy
of the magnon resonance and $T_c$ \cite{mysy}, as observed in neutron scattering experiments \cite{gre}.

\subsection{Acknowledgments}
It is a pleasure to thank Z. B. Su and L. Yu for the joy of a long collaboration on this project, and F. Ye, M. Gambaccini and G. Bighin for their crucial contributions. Partial support of the Cariparo Foundation(Excellence Project: Macroscopic Quantum Properties of
Ultracold Atoms under Optical Confinement") is gratefully acknowledged.

\end{document}